\documentclass[%
 reprint,
 aps, physrev
]{revtex4-2}

\usepackage[T1]{fontenc}
\usepackage[utf8]{inputenc}
\usepackage{graphicx}
\usepackage{tensor}
\usepackage{amsmath}
\usepackage{mathtools}
\usepackage{amsfonts}
\usepackage{hyperref}
\usepackage{wasysym} 
\usepackage{braket}
\usepackage[dvipsnames]{xcolor}
\input{macros}
\newcommand{\spacetildeg}{\blue{{}^3\tilde{g}}}

\newcommand{\beaa}{\begin{eqnarray}}
\newcommand{\eeal}[1]{\end{eqnarray}\label{#1}}

\newcommand{\checking}[1]{\ptc{checking: #1}}

\newcommand{\hbtn}{\blue{\hat{{3}}}}
\newcommand{\hbtone}{\blue{\hat{{A}}}}
\newcommand{\hbttwo}{\blue{\hat{{B}}}}
\newcommand{\hbtthree}{\blue{\hat{{C}}}}

\newcommand{\hbone}{\blue{\hat{\frak{a}}}}
\newcommand{\hbtwo}{\blue{\hat{\frak{b}}}}
\newcommand{\hbthree}{\blue{\hat{\frak{c}}}}

\newcommand{\canonical}{\blue{\widetilde{\mathring\gamma}}}

\newcommand{\rangleg}{\red{\rangle}}

\newcommand{\muthreeg}{\blue{\mu}}

\newcommand{\zR}{\red{\mathring{R}}}

\newcommand{\scri}{\blue{\mycal I}}

\newcommand{\hbbtone}{\blue{\hat{a}}}
\newcommand{\hbbttwo}{\blue{\hat{b}}}

\newcommand{\Rgammas}{\blue{\mathring R}}

\newcommand{\sone}{\blue{i}}
\newcommand{\stwo}{\blue{j}}

\newcommand{\btone}{\blue{A}}
\newcommand{\bttwo}{\blue{B}}

\usepackage[normalem]{ulem}

\renewcommand{\red}[1]{{#1}}
 \renewcommand{\blue}[1]{{#1}}
\renewcommand{\checking}[1]{}
\renewcommand{\ptcheck}[1]{}
\renewcommand{\ptc}[1]{} 
\renewcommand{\ptcr}[1]{}
\renewcommand{\redc}{}

\renewcommand{\zg}{\red{\mathring{g}}}

\begin{document}

\author{Piotr T. Chru\'{s}ciel}
\email{pchrusciel@cft.edu.pl}
\affiliation{Beijing Institute of Mathematical Sciences and Applications, Huairou, China, and
 Center for Theoretical Physics of the Polish Academy of Sciences, Warsaw, Poland} 
 \homepage{homepage.univie.ac.at/piotr.chrusciel}
\author{Raphaela Wutte}
\email{rwutte@hep.itp.tuwien.ac.at}
\affiliation{Mathematical Sciences and STAG Research Centre, University of
Southampton, Highfield, SO17 1BJ Southampton, United Kingdom}

\title{Positivity of holographic energy}
\begin{abstract}
  We prove positivity of a {\redc weighted} holographic energy for four-dimensional spacetimes with negative cosmological constant whose conformal boundary at infinity is conformally static and admits either spherical sections, or toroidal sections with compatible spin structure. 
\end{abstract}

\maketitle
 
\tableofcontents

\section{Introduction}

\ptc{change "preprint" to "reprint" in the document header to return to two columns; deactivate ptc and ptcr to remove my notes}

 Existence of  lower bounds for energy-type expressions  has fundamental importance for well-posedness of every theory, and is closely related to the global behaviour of solutions of the equations.  
  
In the context of the AdS/CFT correspondence, a natural notion of energy is the holographic one~\cite{Papadimitriou:2005ii}, see also \cite{Henningson:1998gx,deHaro:2000vlm,Balasubramanian:1999re,Skenderis:2000in,Cheng:2005wk}, 
which {can be} defined for spacetimes that locally approach Anti-de Sitter at infinity. Such spacetimes have a timelike conformal boundary, $\scri$.
Given a vector field $X$ on $\scri$,   a section $S$ 
of $\scri$, and a metric $g$ 
as made precise below,
we write   $ {Q[S,X]}(g)$ for the holographic charge of $g$ defined as  
\begin{equation}\label{25I26.2}
   {Q[S,X]}(g) = 
  -
   \int_S t^A{}_B X^B ds_A
  \,,
\end{equation}
where $t^A{}_B$ is the  
holographic energy-momentum tensor of $g$. 
The holographic energy is obtained by choosing $X=\partial_t$. This energy is reasonably well understood when the    metric at  $\scri$ is  
 conformal to an ultrastatic metric with Einstein space-sections,  
 in which case the
 holographic energy essentially coincides \cite[Equation~(17)]{Skenderis:2000in}
 with the more usual hyperbolic energy~\cite{ChAIHP,AshtekarDas,HT,AbbottDeser}. For these last metrics several positivity and rigidity theorems are available~\cite{ChHerzlich,Wang,ChGPotaux,ChDelayHPETv1,%
ChGallowayHPET,%
Rallabhandi:2025kti,YiYueHirsch,HuangJangMartin,Hirsch:2025usn,%
BrendleHung,BrendleHung2,GibbonsGPI,GHHP,LeeNeves,GallowayTsang}, 
but nothing has been known so far beyond these cases.
\checking{no logs here whatever the dimension, see Section~\ref{s21III26.1a}}

The aim of this letter is to point out that 
{\redc a suitably weighted 
  holographic energy (see \eqref{25I26.3} below)
  } 
  is positive
for all four-dimensional solutions $(\mcM,g)$ of the Einstein equations with a negative cosmological constant,  
with a conformally static conformal infinity 
\footnote{
A construction of large classes of static vacuum solutions can be found in~\cite{ACD2}. A class of negative-mass solutions (which do not satisfy the hypotheses of our positivity theorem here) can be found in~\cite{ChDelayGluing,CDW}.
},
and  
with sources satisfying the dominant energy condition. Furthermore we either assume  
spherical sections of $\scri$, or   toroidal sections of $\scri$ with trivial induced spin structure on conformal infinity.  
Finally, we suppose that $\mcM$ contains a complete spacelike hypersurface 
$\hyp$
either without boundary
\footnote{Note that this is compatible with any number of asymptotic ends, where the asymptotic behaviour  is only restricted by the requirement of completeness of the metric induced on $\hyp$},
or with a compact boundary. If non-empty, each  component  of the boundary should be   either 
outer  trapped or marginally outer
trapped. 
 
The proof, inspired by~\cite{Cheng:2005wk}, consists of showing how to adapt the Witten argument to such metrics.

Recall that the flagship result in this context, going back to~\cite{GHHP}, is positivity of energy for initial data sets which asymptote to a background with asymptotic imaginary Killing spinors
and which have Birmingham-Kottler asymptotics~\cite{Birmingham}
\footnote{In the physics literature, imaginary Killing spinors are often referred to simply as Killing 
spinors.}
(cf.~\cite{ChHerzlich,Wang}; more recent developments can be found 
in~\cite{ChOberwolfachH,ChGallowayHPET} and references therein). Further milestones include proofs of negative bounds from below for solutions  with toroidal~\cite{BrendleHung,BrendleHung2}, or higher genus~\cite{LeeNeves}, sections of conformal infinity. 

\section{The proof}
 
We consider
four-dimensional 
spacetimes $(\mcM,g)$ solving the Einstein equations with a negative cosmological constant, and with matter fields satisfying the usual positivity conditions. 
We suppose existence of a conformal completion \emph{\`a la} Penrose, with a smooth conformal metric at the conformal boundary.  
When the matter fields decay sufficiently fast 
\footnote{{In the context of the AdS/CFT correspondence, the case of matter fields which do not decay at infinity is also of interest, see \cite{Freedman:2003ax, Boucher:1984yx, Townsend:1984iu} for a discussion of the Witten boundary integral in this context. We do not consider this case here.}}, under mild supplementary asymptotic conditions the arguments in~\cite{Fefferman:2007rka,graham:lee} show that the spacetime metric   $g$   can be written in the form 
 \begin{equation}\label{9I25.11}
   g = x^{-2}
   \big(
    dx^2 + 
    \big(\zgamma_{\btone \bttwo} + x^2 \gammaextwo_{\btone \bttwo}
    \red{+ x^3 \gammaexthree_{\btone \bttwo} 
    +O(x^4)
    }
    \big) dy^\btone dy^\bttwo
     \big)
   \,,
 \end{equation}
 with  a smooth Lorentzian $3$-dimensional metric  $\zgamma_{\btone \bttwo}$ and  smooth tensor fields $\gammaextwo_{\btone \bttwo}$, $\gammaexthree_{\btone \bttwo}$,
   all satisfying  $\partial_x \zgamma_{\btone \bttwo}=0 =\partial_x\gammaextwo_{\btone \bttwo}= \partial_x\gammaexthree_{\btone \bttwo}$.
 Here 
 $$\scri:=\{x=0\}
 $$
 is the conformal boundary at infinity, 
  and
$$
 (y^\btone, x)  \equiv (y^0,y^a, x) \equiv (t,y^a, x)
 $$
  are local coordinates near $\scri$. We assume that the time-coordinate $t$ is globally defined and we set  
 \begin{equation}\label{24I26.1}
   \hyp=\{t=0\}
   \,.
 \end{equation}

  Recall that in the presence of a cosmological constant $\Lambda = -3$ the Witten equation reads 
\begin{equation}\label{28I25.2}
  \gamma^\stwo \hat \nabla_\stwo \psi =0\,,
\end{equation}
where $\stwo$ runs over $\{1, 2, 3\}$, 
with
   $$
    \hat \nabla_\stwo \psi = \nabla_\stwo \psi +\frac i 2 \gamma_\stwo \psi
    \,.
   $$
Recall the Schr\"odinger-Lichnerowicz-Sen-Witten (SLSW) identity, for $\epsilon>0$,  
\begin{align}\label{3V25.1}
    &
        \int_{ {\hyp}\setminus\{x\le \epsilon\}} \left( {|}\hat\nabla \psi   {|}^2  +\langle \psi, (\rho
+ J^\sone \gamma_\sone \gamma_0) \psi\rangle 
 -  {|}\gamma^\stwo \hat \nabla_\stwo \psi  {|}^2  
 \right)  {d \muthreeg} \nonumber \\
     &\quad=
 \Re \int_{\partial\big(\hyp \setminus \{x \le \epsilon\}\big)}
B^\sone[\psi] dS_\sone\,,
\end{align} 
where   $d\muthreeg$ is the metric measure on $\red{\hyp}$, and 
where  $\rho$ is the matter density, $J^i$ the matter current, with the boundary integrand given by
\begin{equation}
  B^\sone[\psi] = \langle {\hat \nabla}^\sone \psi
+ \gamma^\sone \gamma^\stwo \hat \nabla_\stwo \psi, \psi \rangleg\,.
\end{equation} 
Here 
$$
 \langle\psi,   \blue{\phi} \rangle := \psi^\dagger \phi  
 \,,%
$$
where $\psi^\dagger$ denotes complex conjugation and transposition.
We use the convention that 
$\{\gamma^\mu,\gamma^\nu\} = - 2 g^{\mu\nu}$ with a Hermitian $\gamma^0$ and anti-Hermitian $\gamma^i$'s,
and
  \begin{equation}\label{11I26.p1}
    dS_i = \sqrt{\det g_{k\ell}}
     \, \partial_i \, \rfloor \,
     (dx^1\wedge dx^2\wedge dx^3
     )
    \,.
  \end{equation}

 Now,  when the matter fields satisfy the dominant energy condition,   it follows from \eqref{3V25.1}  with  a spinor field satisfying \eqref{28I25.2} that  
  \begin{equation}\label{28I25.3}
%
 \Re
\int_{\partial \red{\hyp}} 
B^\sone[\psi] \,dS_\sone
    \ge \int_{ \red{\hyp}}  |\hat \nabla \psi|^2 d\mu
    \,.
  \end{equation}
So, a finite boundary integral in \eqref{3V25.1} implies that  $|\hat\nabla \psi|$ 
    is square-integrable. Conversely, square-integrability of  $|\hat\nabla \psi|$ and a finite contribution from the matter-fields volume-integral in  \eqref{3V25.1} guarantees a finite boundary integral.
     
     Letting the coordinates $(x,y^A)$ near $\partial \red{\hyp}$ range over $[0,x_0]\times \partial \red{\hyp}$ we have 
  \begin{equation}\label{28I25.4}
    \int_{ \red{\hyp}} |\hat \nabla \psi|^2 d\mu
    \ge \int_{x=0}^{x_0} 
     \int_{\partial \red{\hyp}} |\hat \nabla \psi|^2 
    x^{-3}\sqrt{\det \red{\spacetildeg }} \,
   dx\, dy^1  dy^{2}
    \,, 
  \end{equation}
  where $\red{\spacetildeg }$ is the metric  induced on the level sets of $t$
   by the unphysical spacetime metric
   $$
    \tilde{g} = x^2 g
    \,.
    $$
   The hypothesis of finiteness of the left-hand side of \eqref{28I25.3} implies that $|\hat \nabla \psi|$ must decay faster than $x$. In
fact, for spinor fields  $\psi$ with the asymptotics below we must have
    \begin{equation}
      \label{28I25.92}
       |\hat  \nabla  \psi|  = 
        \blue{O(x^{3/2})} 
        \,,
     \end{equation}
which is then necessary and sufficient for a finite Witten boundary-integral, i.e.\ the boundary integral in \eqref{3V25.1}.

In order to evaluate the above explicitly we use an orthonormal coframe  of the form 
$\theta^{\hat 3} = x^{-1}dx$
and
 \begin{align}\label{14IX25.21}
   &\theta^{\hbtone} = x^{-1}
   \big(
    \ztheta^{\hbtone} + \frac 12\zgamma^{{\hbtone}{\hbttwo}} 
      \big( x^2 \gammaextwo {}_{{\hbttwo}{\hbtthree}}
       +  x^3 \gammaexthree {}_{{\hbttwo}{\hbtthree}}  
       + 
    O(x^4)
    \big) 
     \ztheta^{\hbtthree} 
    \big)
    \,, \nonumber \\
    \
     & \hat A\in \{0,1,2\} 
    \,,
 \end{align}
 where $\{\ztheta^\hbtone\} $ is an ON-coframe for $\zgamma_{\btone \bttwo}$  which is independent of $x$, and we 
put hats over tetrad indices.  
The dual frame takes the form
 \begin{equation}\label{9I25.12BB}
   e_\hbtn = x\, \partial_x
   \,,
   \quad
   e_\hbtone = x 
   \big(
    \ze_\hbtone + x^2 f_\hbtone
    \big)
   \,,
 \end{equation}
 where $\{\ze_\hbtone\}$ is dual to $\{\ztheta^\hbtone\}$, and where the $f_\hbtone$'s are smooth 
 on the conformally completed manifold.   
 We have
 \checking{this is most likely irrelevant, but how do we know that the gamma matrices are constant in this frame?}
 \begin{equation}
 f_\hbtone = -\frac{1}{2} 
 \big (\gammaextwo_{{\hbtone}{\hbttwo}} +
  x \gammaexthree_{{\hbtone}{\hbttwo}} +
  O(x^2)
 \big)
 \zgamma^{{\hbttwo}{\hbtthree}} \ze_\hbtthree
 \end{equation}
The connection coefficients can be written as
  \ptc{there could be log terms in 4 d for static metrics, but they do not affect the end result; see Appendix~\ref{s21III26.1c}} 
\checking{the VAB term is ok, see \eqref{26III26.70}; but so is  CAB, see \eqref{26III26.727}} 
 \begin{align}\label{9I25.12a}
  \displaystyle  \omega_{\hbtone \hbtn } &=  
   -
   \eta_{\hbtone \hbttwo} \theta^\hbttwo + x^{2 } 
   \underbrace{
    V_{\hbtone \hbttwo}
    }_{\gammaextwo_{\hbtone \hbttwo} 
    \red{+  \frac{3}{2} x \gammaexthree_{\hbtone \hbttwo}}
    +\red{O(x^2)}} \theta^\hbttwo
   \,,
   \\ 
   \omega _{\hbtone \hbttwo} &= \zomega_{\hbtone \hbttwo }  + x^{2} 
   \big(
    \underbrace{
    C_{\hbtone \hbttwo}
    }_{ \red{O(x^2)} }  \theta^\hbtn 
   +   
    C_{\hbtone \hbttwo \hbtthree} \theta^\hbtthree 
    \big)
   \,, 
   \label{9I25.12CC}
 \end{align}
 with 
 \begin{equation}
 C_{\hbtone \hbttwo \hbtthree}
    =
      \red{x \zmcD_{[\hbttwo} \gammaextwo_{\hbtone]\hbtthree} }  + \blue{x^2 \zmcD_{[\hbttwo} \gammaexthree_{\hbtone]\hbtthree}}\red{+ \red{O(x^3)}}\,,
 \end{equation}
 where the $ \zomega_{\hbtone \hbttwo }$ are connection one-forms associated with the frame $\ztheta^\hbtone $, with
  \begin{equation}
    \label{gamma2}
    \gammaextwo_{\hbtone \hbttwo} = 
    - 
    \left( \zR_{\hbtone \hbttwo} - \frac{\zR}{4} \zgamma_{\hbtone \hbttwo} \right)
     + o(1)
     \,,
  \end{equation}
where   $\zR_{\hbtone \hbttwo}$ is the Ricci tensor of $\zgamma_{\hbtone \hbttwo}$,  where $o(1)$ is meant as $x\to 0$.

Following~\cite{Cheng:2005wk} we start with a formal solution $\psi$ of  \eqref{28I25.2} which, in a spin frame associated with \eqref{9I25.12BB}, is assumed to have an asymptotic expansion of the form 
      \begin{equation}\label{9I25.4}
        \psi(x, y^A) =  x^{-1/2}
        \big(\zpsi(y^A) + x \,\psibeta(x,y^A)
        \big ) 
        \,\blue{,}
      \end{equation}
where $\psibeta$ is a
bounded polyhomogeneous spinor field on the conformally completed manifold.
Using \eqref{9I25.12a}-\eqref{9I25.12CC} we find 
 \ptcheck{21III, with RW}
 \checking{our equations give O 3/2, which is not enough for integrability in dim 4; \eqref{23III26.63} works but \eqref{23III26.2344} might have a problem, but this is settled in Appendix~\ref{ss27III26.1}}
      \begin{align}
        \hat \nabla_\hbtn  \psi \equiv
         & 
         \
          \nabla_{e_\hbtn}\psi + \frac i2 \gamma_\hbtn  \psi  
          \nonumber 
\\
      = & \ 
      x^{\red{3/2}} \partial_x\psibeta
       -\frac 12 x^{-1/2} (1 - i \gamma_\hbtn)  \zpsi \nonumber \\
       & \
       -\frac{x^{3/2}}{4} C_{\hbtone \hbttwo} \gamma^{\hbtone} \gamma^{\hbttwo} \zpsi
       + \frac i2 \gamma_\hbtn   x^{\red{1/2}} \psibeta   
          \nonumber 
\\
 & \   
       +
       \frac 12  x^{\red{1/2}}  \psibeta     -\frac{x^{\frac 52 }}{4} C_{\hbtone \hbttwo} \gamma^{\hbtone} \gamma^{\hbttwo} \psibeta
      \,,
      \label{9I25.4a}
 \\
      \hat \nabla_\hbone \psi 
       \equiv
       &
       \
         \nabla_{e_\hbone}\psi + \frac i2 \gamma_\hbone \psi  
          \nonumber  
\\
          =
            & \ 
          x^{1/2}\zmcD_{\ze_\hbone} \zpsi 
          + x^{\red{3/2}} \zmcD_{\ze_\hbone} \psibeta
          +  x^{5/2} \zmcD_{f_\hbone} \zpsi
          + x^{\red{\frac 72} } \zmcD_{f_\hbone}\psibeta 
          \nonumber 
\\
          & \ 
          + \frac { x^{-1/2}}2 i \gamma_\hbone 
          \big( 1
         - i    \gamma^{\hbtn }\big)\zpsi
          + \frac{x^{\red{1/2}}}{2}  \gamma_{\hbone}
          (  \gamma^{\hbtn} + i )\psibeta
          \nonumber 
\\
          & \ 
          - x^{3/2} \big( \frac{V_{\hbone \hbttwo}}{2} \gamma^{[\hbttwo} \gamma^{\hbtn]} 
          +
           \frac{C_{\hbttwo \hbtthree \hbone }}{4} \gamma^{[\hbttwo} \gamma^{\hbtthree]} \big) \zpsi \nonumber 
\\
          & \  - x^{\frac 52 }   \big( \frac{V_{\hbone \hbttwo}}{2} \gamma^{[\hbttwo} \gamma^{\hbtn]} + \frac{C_{\hbttwo \hbtthree \hbone }}{4} \gamma^{[\hbttwo} \gamma^{\hbtthree]} \big) \psibeta 
          \,,
      \label{9I25.4b}
        \end{align}
      where $\zmcD$ is the spinor covariant derivative associated with the metric $\zgamma_{\btone \bttwo}dy^\btone dy^\bttwo$.
        
   Multiplying \eqref{9I25.4a} by $x^{1/2}$ and passing in the resulting equation to the limit $x\to 0 $, 
       the fall-off requirement \eqref{28I25.92}   
       provides a  condition  already pointed out in~\cite{Cheng:2005wk}: 
 \ptcheck{21III, with RW}
     \begin{align}\label{28I25.5}
      \boxed{(1 - i \gamma_{\hbtn})  \zpsi = 0 } 
        \,.
     \end{align}
     From now on we assume that \eqref{28I25.5} holds. 

Let us set
\begin{equation}
 \psionehalf := \psibeta\vert_{x=0}
 \,.
\end{equation}
Demanding that the terms of  order $x^{1/2}$ in \eqref{9I25.4a} vanish 
       we get
 \ptcheck{21III, with RW}
      \begin{equation}
        \label{psione}
      (1+ i \gamma_{\hbtn}) \psionehalf  = 0\,.
      \end{equation}
    From \eqref{9I25.4b} at order $x^{1/2}$ we have
 \ptcheck{21III, with RW}
     \begin{equation}
    0=  \zmcD_{\ze_\hbone} \zpsi  + \frac{i}{2}  \gamma_{\hbone}
          ( 1 - i \gamma^{\hbtn} ) \psionehalf  
          = \zmcD_{\ze_\hbone} \zpsi
          +  i \gamma_{\hbone}
          \psionehalf   
          \,,
          \label{24I26.11}
     \end{equation}
     where we used \eqref{psione}. 
     
Note that if $\psibeta$ vanishes at $\{x=0\}$ we obtain
     \begin{equation}
    \zmcD_{\ze_\hbone} \zpsi  = 0\,.
     \end{equation}
     It then follows that $\zgamma$ is flat,  a case which has already been covered elsewhere~\cite{Maerten,CMT,BrendleHung,BrendleHung2}.
      
Multiplying \eqref{24I26.11} with $\gamma^{\hbone}$ we get   
 \ptcheck{21III, with RW}
\begin{equation}
  \boxed{
\psionehalf  = - \frac{i}{2}  \gamma^{\hbone} \zmcD_{\ze_\hbone} \zpsi}
 \,,
\label{psi1sol}
\end{equation}
as already pointed out in \cite[Equation (5.15)]{Cheng:2005wk}.
Plugging \eqref{psi1sol} back into \eqref{24I26.11} yields 
\begin{equation}
  \label{eqspin}
\Big( \delta_{\hbone}^{~\hbtwo} 
 + \frac{1}{2}  \gamma_{\hbone}  \gamma^{\hbtwo}    
  \Big)
    \zmcD_{\ze_\hbtwo} \zpsi = 0 
  \quad
  \Longleftrightarrow
  \quad  
   \boxed{
    \gamma^{\hbtwo} \gamma^{\hbone}  \zmcD_{\ze_\hbtwo} \zpsi
    = 0
    }
  \,.
\end{equation}

We are ready now to prove our claim. Suppose that the conformal metric at $\scri$ contains a static metric in its class. We can choose a conformal representative which is  {ultrastatic},
\begin{equation}\label{24I26.12}
  \zgamma = -dt^2 + \zgamma_{ab}dx^a dx^b
  \,,
   \quad
   \partial_t \zgamma_{ab} = 0 
  \,.
\end{equation} 
Then
$\zmcD_{\ze_\hbtwo} \zpsi$ is the spinor derivative associated with the metric induced on the level sets of $t$ within $\scri$,  and
 \eqref{eqspin} is the two-dimensional twistor equation.
Since this equation is conformally invariant, the space of solutions of this equation is 
 essentially  
the same for all metrics on a two-dimensional sphere $\mathbb{S}^2$, and for all metrics on 
 $\mathbb{T}^2$ carrying   its
trivial spin structure,
 cf.\ 
\cite[Note A.2.2]{ginoux2009dirac}. 
 
Indeed, for any smooth metric $\zgamma_{ab}$ on a two-dimensional  compact manifold  we can write
(cf., e.g., \cite{MazzeoTaylor})
\begin{equation}\label{16III26.21}
 \zgamma_{ab}=e^{2u} \canonical_{ab}
  \,,
\end{equation}
where $\canonical$ has constant scalar curvature in $\{0,\pm 2\}$, 
for some smooth function $u(x^a)$.
Letting
\begin{equation}\label{15III26.11}
\widetilde{\zpsi} = e^{\frac{  u}{2}} \zpsi
\,,
\end{equation}
we have
\begin{equation}
  \gamma^{\hbbttwo} \gamma^{\hbbtone}    
 \widetilde{\zmcD_{\widetilde e_\hbbttwo}} \, \widetilde{ \zpsi} = 
 e^{{-\frac{u}{2}} }  \gamma^{\hbbttwo} \gamma^{\hbbtone}     \zmcD_{e_\hbbttwo} \zpsi\,.
\end{equation}
For further use we note that if
\begin{equation}\label{2II26.4a}
  X^A = (\zpsi)^\dagger \gamma^{\hat 0} \gamma^A \zpsi
   \,,
\end{equation}
then after the conformal rescaling \eqref{16III26.21} we will have
\begin{equation}\label{2II26.4}
  \widetilde X^A: = 
(\widetilde \zpsi)^\dagger \gamma^{\hat 0} \gamma^A \widetilde\zpsi = e^u X^A
   \,.
\end{equation}

Given a solution of \eqref{eqspin} we can define $\psionehalf $ using \eqref{psi1sol}. In order to satisfy \eqref{24I26.11} we need
\begin{equation}
  \label{eq:161261}
 \psionehalf  =  - i \gamma_{\hat 1} \zmcD_{\ze_{\hat 1} } \zpsi = - i \gamma_{\hat 2} \zmcD_{\ze_{\hat 2} } \zpsi\,,
\end{equation}
which is satisfied when \eqref{eqspin} holds. 
 \checking{works in all dimensions, see Section~\ref{s24III26.1} for dim 3 and Scribble for all dimensions}

We claim, now, that when $\scri$ has spherical topology, then a solution $\psi$ of   \eqref{28I25.2} exists for any asymptotic values of $\zpsi$ and of $\psionehalf$ as just described; in the toroidal case this remains true provided the spin structure on $\red{\hyp}$ induces the trivial spin structure on $\scri\cap \hyp$ 
\footnote{Our argument does not provide any information about  higher genus two-dimensional manifolds, as no non-trivial solutions of \eqref{eqspin} exist there.}.
For this
 we write 
$$
  \psi(x,y^A) =x^{-1/2}\zpsi(y^A)+ x^{1/2}\psionehalf (y^A)
  + \phi(x,y^A) 
  \,,
$$
with $\phi\in L^2(\hyp)$. We thus need
\begin{equation}\label{28I25.2p1}
  \gamma^\stwo \hat \nabla_\stwo\phi = - \gamma^\stwo \hat \nabla_\stwo
   \big(x^{-1/2}\zpsi
  + x^{1/2}\psionehalf 
  \big)
  \,.
\end{equation}
If the right-hand side is in $L^2(\hyp)$, one shows by standard methods that a solution $\phi\in L^2(\hyp)$ exists, with the property that the contribution of $\phi$ to the boundary term vanishes  (cf., e.g., \cite{BartnikChrusciel1,ChHerzlich}).

For any such solution the boundary integral in \eqref{3V25.1} is positive and finite,
\begin{equation}\label{2II26.1}
 0\le  \Re \int_{S} 
  B^\sone[\psi ] 
 dS_\sone < \infty\,,
\end{equation}
where we used $S$ for  $\hyp\cap\scri$.

One can use the results in~\cite{LockhartMcOwen85,Lee:fredholm}
to show that 
 near $\{x=0\}$ 
the solution $\psi$ so obtained takes the form
\begin{equation}\label{24III26.91a}
  \psi (x,y^A)= x^{-1/2} \phi(x,y^A) + x^{5/2} \log x \, \phi_{\log{}} (x,y^A) 
  \,, 
\end{equation}
for some smooth-up-to-boundary fields $\phi$ and $\phi_{\log{}}$,
with
\begin{equation}\label{24III26.91b} 
   \phi_{\log{}} (0,y^A) =  -
     \frac{i}{4} \gammaextwo_{\hbone \hbttwo}
     \gamma ^ \hbone
     \gamma^{\hbttwo} 
   \gamma^{\hbthree} \zmcD_{\ze_\hbthree} \zpsi
  \,.
\end{equation}
In particular $\psi$ 
 has the asymptotic behaviour needed for the calculations 
in~\cite{Cheng:2005wk}
(the log term gives a vanishing contribution to the boundary integral 
because its leading coefficient is orthogonal to $\zpsi$).
It is shown in the last reference that when \eqref{28I25.5}-\eqref{psione} hold the Witten boundary integral,
before passing to the limit $x\to 0$,
is proportional
to
\begin{align}\label{25I26.1a}
   {x^{-1}  \mathcal \Re}\int_{\hyp\cap\{x= \epsilon\}} 
  &
   \sqrt{\det \zgamma_{ab}}\,
  (\zpsi)^\dagger
  \Big[
    (\gamma^\hbone  \zmcD_{\ze_\hbone})^2 
    \nonumber \\ 
     &
     + (\Rgammas_{\hbtwo \hbtone} - \frac{\Rgammas}{4}  \zgamma_{\hbtwo \hbtone} )
    \gamma^{\hbtwo} \gamma^{\hbtone}  
    \Big] \zpsi \nonumber \\
    &
     + Q[\hyp\cap\scri,X](g) + o(1)
    \,,
\end{align}
with 
\begin{equation}\label{2II26.4aa}
  X^A = (\zpsi)^\dagger \gamma^{\hat 0} \gamma^A \zpsi
   \,.
\end{equation}
This seemingly leads to a divergent boundary term, which would contradict \eqref{2II26.1}. However, Equation \eqref{3V25.1} shows that the divergent part of \eqref{25I26.1a} must integrate out to zero; in fact, a commutation calculation shows that the singular part of the integrand vanishes when \eqref{eqspin}  holds.
{\redc 
From what has been said, 
 using \eqref{25I26.1a}
we obtain  
\begin{align}\label{25I26.3}
  \red{Q_{CW}[S,\bar X]}(g) 
  :=
  -
   \int_S t^A{}_B  {\widetilde X^B} e^{-u} dS_A
 \ge 0
  \,,
\end{align}
where $e^{2u}$ is the conformal factor defined in \eqref{16III26.21}.
Note that this factor can be chosen to be equal to one when $\zgamma_{ab}$ has constant scalar curvature, recovering in this case the already-known positivity results.
} 
In case of a spherical $\scri$, letting $\psi$ run over all possible solutions of the equations above
 one concludes, as in \cite{Maerten,CMT,YauAndFriends}, that in the zero-space-momentum frame the energy $m$, center of mass  $\vec c$, and angular momentum $\vec j$ satisfy 
 \checking{justification in Section~\ref{s21III26.2}}
\begin{equation}\label{2II26.7}
m\ge \, \sqrt{|\vec
c|^2+|\vec j|^2+2|\vec c \times \vec j|}
\,,
\end{equation} 
where $\vec c \times
\vec j$ is the vector product, while $|\vec j|=
\sqrt{(j^1)^2+(j^2)^2+(j^3)^2}$, etc. 
For a toroidal $\scri$ with trivial induced spin structure one has instead
\bel{Min2}
  m\ge \, |\vec j|\;, \qquad |\vec j|:=\sqrt{(j^1)^2+(j^2)^2}
 \;;
 \ee
see \cite{CMT} for definitions and details. 
{\redc
Here  the global charges are calculated by using  in  \eqref{25I26.3}  the Killing vectors $  {\widetilde X^A}$ of  Anti-de Sitter  space in the spherical case, or of the quotient thereof in the toroidal case.
}

The question then arises, which metrics saturate the inequality in \eqref{25I26.3} with $\blue{\widetilde X}=\partial_t$.  
Then, 
assuming that the conformal metric at $\scri$ is the same as for Anti-de Sitter, it is shown in~\cite[Corollary~9.4]{Hirsch:2025usn} 
that the initial-data hypersurface $\hyp$ embeds isometrically into   
Anti-de Sitter spacetime. 
More generally, under our assumptions it follows from \eqref{3V25.1},  with $\psi$ satisfying \eqref{28I25.2}, that the right-hand side of \eqref{25I26.3} vanishes  if and only if the matter contribution vanishes  and there exists along $\hyp$ a spinor field satisfying $\hat \nabla_\stwo \psi=0$. 
Deforming $\hyp$ within a development $\mcM$ of the data on $\hyp$, assuming there is one, we find that  $\mcM$ admits an imaginary Killing spinor. 
Theorem~5.1 of \cite{Baum:2004uyz} leads  to the conclusion that the spacetime metric is a Siklos wave~\cite{SiklosL,Leitner2001The} 
(which appears in the list given in~\cite[Theorem~5.1]{Baum:2004uyz} as pp-manifold)  (compare~\cite{leitner2003imaginary, Baum})
\footnote{Every imaginary Killing spinor is a twistor spinor, whence the  relevance of~\cite{Baum:2004uyz}, which is concerned with twistor spinors. 
Theorem~5.1 of \cite{Baum:2004uyz} shows that the metric must be a  Siklos metric, since twistor spinors on  Fefferman metrics listed there are never imaginary Killing spinors~\cite{Leitner2001The}.
}.  
Whether or not globally well-behaved such metrics exist in four spacetime dimensions requires further investigation.

In our positivity claim we assumed a conformally static infinity because this provides a natural zero-energy reference. But, as already pointed out by Cheng and Skenderis~\cite{Cheng:2005wk}, it would be of interest to determine all backgrounds for which the Witten argument 
provides positive charges. It follows from our calculations above that positivity
also
 holds for 
metrics with the same conformal infinity as 
Siklos waves,  
for which the positive charge is associated with a null conformal Killing vector on $\scri$.

To make things precise, recall that the Siklos waves take the form 
\begin{equation}
 x^{-2} (-2 du ds + f(s, x, y) ds^2 + dx^2 +dy^2)\,,
\end{equation}
where 
\begin{equation}
  f(s,x,y)= \zf(s,y) + \frac{x^2}{2} \partial_y^2 \zf(s, y) + x^3\threef (s,y) + ...
  \,,
\end{equation} 
and have an imaginary Killing spinor. 
They induce on $\scri$ the metric 
\begin{equation}
\zgamma = -2 du ds + \zf(s,  y) ds^2  +dy^2 
\label{7II25.1}
\end{equation}
which has vanishing Ricci scalar, with the only non-vanishing component of the Ricci tensor being
$\zR_{ss}=- \frac{1}{2} \partial_y^2 \zf$, and the only non-zero component 
of the Cotton tensor being
 $\mathring C_{ss}= \frac{1}{2}  \partial_y^3 \zf$. 
Hence the boundary metric is  conformally flat if and only if $\partial_y^3 \zf \equiv 0$.
Every Lorentzian metric $g$ with the  conformal structure on $\scri $ given by \eqref{7II25.1}, which satisfies the dominant energy condition and which is vacuum to sufficiently high order, on a manifold which has a partial Cauchy surface $\hyp$  inducing the trivial spin structure on $\hyp\cap\scri$,  has positive charge $Q[\hyp\cap\scri,\partial_u](g)$. A similar statement holds for  higher-dimensional Siklos waves.

Let $\zg$ be static and let $g$ and $\zg$ share the same conformal structure on $\scri$. When both metrics are, say, vacuum it was shown in~\cite{ChWutteEnergy} that the functional
\begin{align}\label{25I26.6}
  {Q[S,\partial_t]}(g) -  {Q[S,\partial_t]}(\zg)
\end{align}
is a Noether charge 
(compare~\cite{Papadimitriou:2005ii,HollandsIshibashiMarolf})  
equal to the Hamiltonian energy of $g$ relative to $\zg$. 
Recall that a perturbative analysis of the positivity of mass goes back to~\cite{AbbottDeser}, where a quadratic expansion of the energy around the Anti-de Sitter metrics confirmed positivity  for nearby metrics; compare~\cite{BCN}. Now, the calculations in that last reference have been carried out for perturbations of any static metrics, in any dimension. In particular the calculations 
there are directly relevant to  static vacuum metrics near the Anti-de Sitter metric, constructed in~\cite{ChDelayKlinger,ACD2}, which are parameterised by conformally static conformal structures. These are
 stationary points of energy, thus have a Taylor expansion in terms of the perturbations of the metric which starts with quadratic terms. Since the second-derivative operator of the energy functional is strictly positive at the Anti-de Sitter metric, and depends continuously upon the metric,  the arguments in~\cite{BCN}, including the possibility of realising a suitable gauge, prove  positivity of Hamiltonian mass %
\footnote{We are grateful to Klaus Kröncke for pointing this out.}
near these metrics, and hence of \eqref{25I26.6}, in all spacetime dimensions $n+1\ge 3$
\footnote{See~\cite{ChWutte, Chrusciel:2024vle} for $n=2$.}.  

In $3+1$-dimensions we hence obtain  
two independent
 inequalities, namely \eqref{25I26.3} with $ {\widetilde X}=\partial_t$
 and 
\begin{align}\label{25I26.6wer}
 \red{Q[S,\partial_t]}(g) \ge  \red{Q[S,\partial_t]}(\zg)
\end{align}
  when a 
  vacuum metric $g$ is close to a vacuum 
  static background $\zg$, both sharing the same induced geometry at $\scri$,  and when $\zg$ itself is close to  the AdS metric. Under these restrictions, equality in \eqref{25I26.6wer} occurs only 
  when   $g=\zg$.

Note that we have been concentrating on spacetime dimension four, because the requirement of convergence, as $\epsilon\to 0$, of the volume integral of \eqref{3V25.1} in spacetime dimension $n+1$ requires $|\nabla\psi|= O(x^{\frac{n-1}{2} +\epsilon})$ for some $\epsilon > 0$, which might impose more conditions on the boundary geometry when $n>3$. It is tempting to conjecture that positivity of $Q_{CW}$ holds in all spacetime dimensions $n+1\ge 4$ for all conformally static $\scri$'s with sections which carry solutions of the  
 $(n-1)$-dimensional twistor  equation:
\begin{equation} 
\Big( \delta_{\hbone}^{ \hbtwo} 
 + \frac{1}{(n-1)}  \gamma_{\hbone}  \gamma^{\hbtwo}    
  \Big)
    \zmcD_{\ze_\hbtwo} \zpsi = 0 \,; 
     \label{eqspinnnew}
\end{equation}
cf.\ e.g.~\cite{BaumEtAl,ginoux2009dirac}. 
The fact that the conjecture is true in $(3+1)$-dimensions has been shown above; it
 is true in $(4+1)$-dimensions by calculations identical to the above together with the last paragraph of \cite[Section 6]{Cheng:2005wk}.

\section{Conclusions}
In this letter  we established positivity of
{\redc the  weighted holographic energy \eqref{25I26.3}
}
for four-dimensional spacetimes 
with or without black hole boundaries, with conformally
 static infinity,
 with spherical cross-sections of infinity, or with toroidal cross-sections with a compatible spin structure.
 We further established positivity of a holographic charge associated with a null conformal Killing vector for conformal boundary geometries induced by a Siklos wave in all dimensions.
  Our argument exploits 
 existence of solutions to the twistor equation on sections of the conformal boundary.
The same method applies to prove positivity for all $(4+1)$-dimensional spacetimes  which possess a conformally static boundary geometry at infinity with sections which admit solutions of the three-dimensional twistor equation. We conjecture that a similar statement holds true in all spacetime dimension greater than five.

{\redc 
Since
\begin{equation}\label{20III.1}
   \red{Q_{CW}[S,\bar X]}(g) \equiv  \red{Q [S,e^{-u}\bar X]}(g) 
   \,,
\end{equation} 
one can instead think of $Q_{CW}$ as the usual holographic charge with respect to a rescaled vector field. Now,  the divergence theorem shows that $ Q [S,\partial_t](g) $ is independent of $S$  when $\partial_t$ is a conformal Killing vector of the boundary geometry because $t_{AB}$ is transverse and traceless in odd space dimensions.   
But $\red{Q_{CW}[S,\partial_t]}(g) $ will depend upon $S$ in general because $e^{-u}\partial_t$ is not a conformal Killing vector  
 on $\scri$ unless $u$ is constant.
This implies that  $ Q_{CW}  $    can be radiated away, which can be considered as a desirable feature, with positivity providing an upper bound on the amount of charge that can be emitted. 
 
Our result should be contrasted with the analysis in~\cite{HicklingWiseman}, where it is proved
 that $(3+1)$-dimensional \emph{static} vacuum conformally 
   smooth metrics with conformally compact static slices
 without boundary have \emph{negative or zero holographic} energy; recall that many such metrics exist near Anti-de Sitter spacetime by~\cite{ACD, ACD2}. 
 It is rather counterintuitive that the introduction of the conformal factor arising from the solutions of the twistor equation changes the sign of the charge integrals, as follows from our analysis. This implies in particular that the integrand of \eqref{25I26.2} can  have a constant sign for 
 such
 metrics only if it vanishes.
}

\begin{acknowledgments}
We are grateful to Pawe\l\ Nurowski, Kostas Skenderis, Paul Tod and Yiyue Zhang for useful discussions, and to Thomas Leistner, Felipe Leitner,
Robin Graham and Jaros\l aw Kopi\'nski  for bibliographical advice.  
 RW acknowledges
support from the STFC consolidated grant ST/X000583/1 “New Frontiers in Particle Physics,
Cosmology and Gravity”, the Heising-Simons Foundation under the “Observational Signatures of Quantum Gravity” collaboration grant 2021-2818 and the U.S. Department
of Energy, Office of High Energy Physics, under award DE-SC0019470. 
We thank the Erwin Schrödinger Institute for Mathematics
and Physics, Vienna, for hospitality where part of this work was carried out.
PTC research was further supported in part by the NSF under Grant No.
DMS-1928930 while he was in residence at the Simons Laufer Mathematical
Sciences Institute (formerly MSRI) in Berkeley during the Fall 2024 semester.
\end{acknowledgments}

%
%

  \bibliographystyle{amsplain}

\bibliography{ChruscielWutteLetter-minimal, Chrusciel-Wutte-spinors} 
\end{document}